\documentclass[useAMS,usenatbib]{mn2e}
\usepackage[a4paper,margin=2cm]{geometry}
\usepackage{times}
\usepackage{epsfig}
\usepackage[]{graphicx} 
\usepackage{verbatim} 
\usepackage[fleqn]{amsmath}
\usepackage{amssymb}
\usepackage{calc}
\usepackage{caption}
\usepackage[normal]{threeparttable}
\newcommand{\comments}[1]{} 
\graphicspath{{./plots/}}

\title[Broad, weak 21\,cm absorption in an early type galaxy]{Broad,
  weak 21\,cm absorption in an early type galaxy: spectral line
  finding and parametrization for future surveys}
\author[J.~R. Allison et al.]{J.~R. Allison$^{1}$\thanks{E-mail:
    jra@physics.usyd.edu.au}, S.~J. Curran$^{1,2}$,
  E.~M. Sadler$^{1,2}$ and S.~N. Reeves$^{1,2,3}$ \\$^{1}$Sydney
  Institute for Astronomy, School of Physics A28, University of
  Sydney, NSW 2006, Australia\\$^{2}$ARC Centre of Excellence for
  All-sky Astrophysics (CAASTRO)\\$^{3}$CSIRO Astronomy \& Space
  Science, P.O. Box 76, Epping NSW 1710, Australia}
\begin{document}

\date{}

\pagerange{\pageref{firstpage}--\pageref{lastpage}} \pubyear{2012}

\maketitle

\label{firstpage}

\begin{abstract}
  We report conclusive verification of the detection of associated
  \mbox{H\,{\sc i}} 21\,cm absorption in the early-type host galaxy of
  the compact radio source PMN\,J2054--4242. We estimate an effective
  spectral line velocity width of $418 \pm 20$\,km\,s$^{-1}$ and
  observed peak optical depth of $2.5 \pm 0.2$\,per\,cent, making this
  one of the broadest and weakest 21\,cm absorption lines yet
  detected. For $T_\mathrm{spin}/f > 100$\,K the atomic neutral
  hydrogen column density is $N_\mathrm{HI} \gtrsim 2 \times
  10^{21}$\,cm$^{-2}$. The observed spectral line profile is
  redshifted by $187 \pm 46$\,km\,s$^{-1}$, with respect to the
  optical spectroscopic measurement, perhaps indicating that the
  \mbox{H\,{\sc i}} gas is infalling towards the central active
  galactic nucleus. Our initial tentative detection would likely have
  been dismissed by visual inspection, and hence its verification here
  is an excellent test of our spectral line detection technique,
  currently under development in anticipation of future
  next-generation 21\,cm absorption-line surveys.
\end{abstract}


\section{Introduction}\label{section:introduction}

\begin{figure*}
\centering
\includegraphics[width = 2.0\columnwidth]{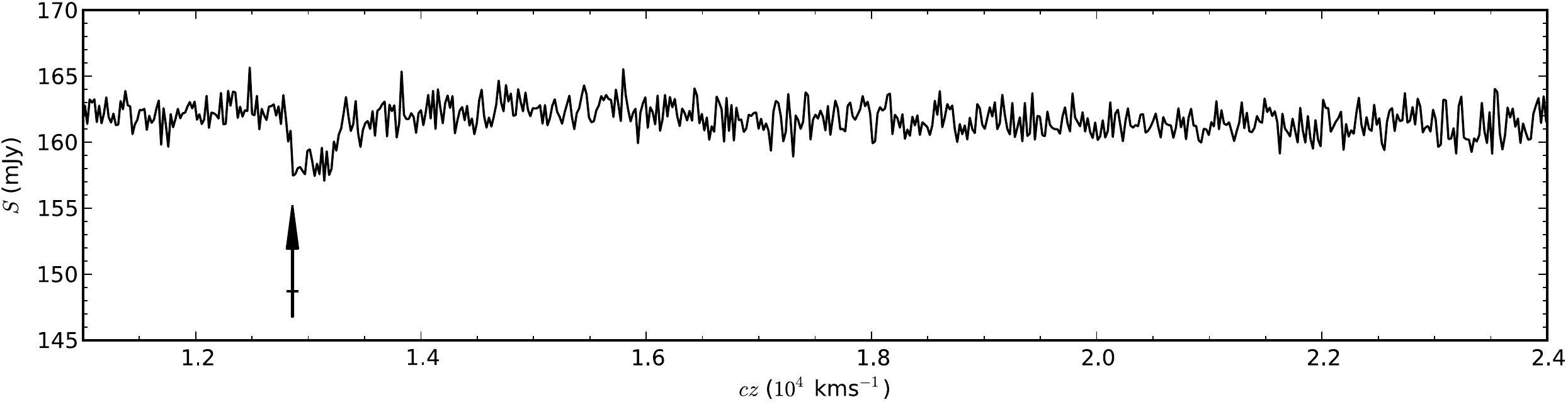}
\caption{The synthesized-beam weighted spectrum, combining ATCA 21\,cm
  data from observations in 2011 April and 2012 June. The data were
  binned to a resolution of 21\,km\,s$^{-1}$, which is approximately
  three times that of the individual spectra. The optical
  spectroscopic redshift is indicated by the vertical arrow
  \citep[$z_\mathrm{opt} = 0.042893\pm0.000150$;][]{Jones:2009}.}
\label{figure:atca_spectrum}
\end{figure*}

The cold atomic neutral hydrogen content of galaxies (\mbox{H\,{\sc
    i}}) plays a vital role in the evolution of galaxies, through
formation of stars and efficient accretion of matter on to active
galactic nuclei (AGNs). By using the 21\,cm hyperfine transition of
\mbox{H\,{\sc i}}, observed as absorption against a background radio
source of known brightness, we can measure the distribution and
kinematics of line-of-sight cold gas ($T \sim 100$\,K). Previous
observations of the 21\,cm line in associated absorption often exhibit
spectral line profiles with both a deep, narrow component
($\sim$\,10\,km\,s$^{-1}$) and broad wings from a shallow component
($\sim$\,100\,km\,s$^{-1}$) \citep[e.g.][]{Mirabel:1989,
  Morganti:2001, Morganti:2005a, Salter:2010, Allison:2012a}. From
observations at high spatial resolution there, is evidence that in
some cases these broad, shallow components might be associated with a
circumnuclear torus \citep[e.g.][]{Struve:2010,
  Morganti:2011}. Additionally, \cite{Morganti:2005b} have
successfully detected extremely broad absorption within gas-rich
hosts, each of which has an existing deep, narrow absorption-line
component. These systems are associated with high-velocity gas
outflows, providing a probe of the feedback between the central radio
source and the star-forming interstellar medium of its host
galaxy. Other examples of extreme broadening include the detection of
\mbox{H\,{\sc i}} absorption towards the candidate binary black-hole
4C\,37.11 \citep{Maness:2004,Morganti:2009a}, with a full width at
zero-intensity of 1600\,km\,s$^{-1}$ and dominated by two narrow, deep
peaks. However, there are currently very few detections that are
dominated by broad absorption, with no (or very little) accompanying
narrow component \citep[see e.g.][]{Jaffe:1990}.

In previous work, \citet{Allison:2012a} reported the tentative
detection of broad ($\Delta{v_\mathrm{FWHM}} \sim 500$\,km\,s$^{-1}$)
\mbox{H\,{\sc i}} 21\,cm absorption towards the compact flat-spectrum
radio source, PMN\,J2054--4242 \citep{Griffith:1993,Healey:2007},
using a spectral line detection method based on a Monte Carlo Bayesian
technique. From observations using the Australia Telescope Compact
Array Broadband Backend \citep[CABB;][]{Wilson:2011}, and by
considering the ratio of the marginal likelihoods, the data were found
to warrant the inclusion of the spectral line hypothesis over the
continuum-only hypothesis. Here we report further observations to
conclusively verify the presence of this broad and shallow absorption
line. This is one of the broadest and weakest 21\,cm absorption lines
yet detected in isolation, with no previous indication of absorption
from an existing deep, narrow component. Conclusive verification of
this detection provides an excellent test of our spectral line finding
method, a technique that can be used for detecting and parametrizing
such spectral lines in future large-scale absorption surveys.

Throughout this Paper we adopt a flat $\Lambda$CDM cosmology with
$\Omega_\mathrm{M}$ = 0.27, $\Omega_{\Lambda}$ = 0.73 and $H_{0}$ =
71\,kms$^{-1}$\,Mpc$^{-1}$. Quantities estimated from the data are
given by their mean and 1\,$\sigma$ uncertainty unless otherwise
stated.

\section{Observations and data reduction}

In addition to the observations in 2011 April reported by
\cite{Allison:2012a}, a further 12\,h observation of PMN\,J2054--4242,
at $\rmn{RA}(\rmn{J}2000) = 20^{\rmn{h}} 54^{\rmn{m}} 01\fs79$ and
$\rmn{Dec.}(\rmn{J}2000) = -42\degr 42\arcmin 38\farcs7$, was carried
out in 2012 June 17--18. The CABB system provides a zoom band of
64\,MHz across 2048 channels, which at the centre frequency of
1.342\,GHz has a velocity resolution $\sim7$\,km\,s$^{-1}$. The
six-element ATCA was arranged in the 6D East-West configuration, with
baseline distances in the range 0.08 -- 5.88\,km. With this
configuration we obtained an angular scale sensitivity range of
approximately 8 -- 600\,arcsec, and a primary beam full width at
half-maximum (FWHM) of 35\,arcmin. We observed the target source in
20\,min scans, which were interleaved with 1.5\,min observations of
the nearby secondary calibrator, PKS\,B2211--388
($S_\mathrm{1.384\,GHz} = 1.86\pm0.01$\,Jy),
\footnote{http://www.narrabri.atnf.csiro.au/calibrators/\label{footnote:atca_cals}}
for gain and band-pass calibration. For the calibration of the
absolute flux scale, we observed PKS\,B1934--638
($S_\mathrm{1.384\,GHz} = 14.94\pm0.01$\,Jy)\footnotemark[1] at
regular intervals separated by 2\,h. For the 2012 June epoch, we
achieved a total integration time of 8.6\,h on the target source.

We implemented the same data reduction procedure as reported by
\cite{Allison:2012a}, again subtracting the baseline ripple signal
from other continuum sources within the field of view. The data were
flagged, calibrated and imaged using tasks from the
\textsc{Miriad}\footnote{http://www.atnf.csiro.au/computing/software/miriad/}
package \citep{Sault:1995}, and implemented using a purpose-built CABB
\mbox{H\,{\sc i}} data reduction pipeline. We updated our original
procedure to include some minor improvements to the calibration and
baseline ripple subtraction, and have applied these here to the data
obtained both in 2011 April and 2012 June. Assuming that the flux
density of PMN\,J2054--4242 has not varied significantly between the
two epochs (which was not evident from our observations), we
constructed a single, natural weighted data cube from the combined
visibility data. Since the ATCA does not observe at constant radial
velocity (known as Doppler tracking), the observed frequency channels
for each epoch correspond to slightly different velocities. To obtain
approximately uniform variance in flux across the velocity spectrum
(while still retaining spectral information), we binned the combined
data to a resolution of 21\,km\,s$^{-1}$, which is approximately three
times that of the individual spectra. Fig.\,\ref{figure:atca_spectrum}
shows the resulting synthesized-beam weighted spectrum centred on
PMN\,J2054--4242. Based on the median absolute deviation from the
median \citep[e.g.][]{Whiting:2012}, the estimated per
spectral-channel uncertainty is 1.05\,mJy, which is consistent with
the uncertainties in the spectra for each epoch and the subsequent
velocity binning (see Table\,\ref{table:hi_summary}). The 21\,cm
\mbox{H\,{\sc i}} absorption can clearly be seen upon visual
inspection, slightly redshifted with respect to the systemic velocity,
given by optical spectroscopy from the 6dF Galaxy Survey
\citep[6dFGS;][]{Jones:2009}.  This feature is seen throughout
observations in both epochs, and in both polarizations, strongly
indicating that it is not generated by a transient artefact such as
radio frequency interference. We also do not see this feature in
spectra that are randomly selected at positions away from the
source. At a third epoch, during another observing program in 2012
April, we obtained further low spectral resolution (230\,km\,s$^{-1}$)
CABB data. While not useful for parametrizing the absorption, the
21\,cm spectral line profile is wide enough to be seen across three
channels in the data and so provided additional verification of our
detection.

\section{Analysis}\label{section:analysis}

\begin{figure*}
\centering
\includegraphics[width = 1.0\textwidth]{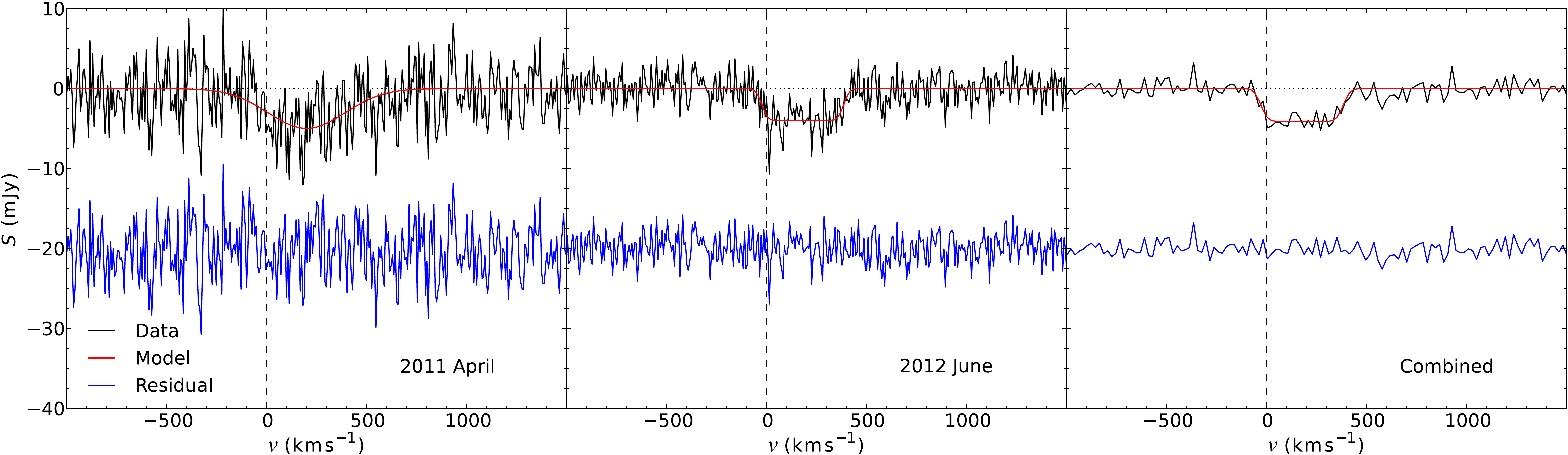}
\caption{Results of fitting to spectra from 21\,cm observations in
  2011 April and 2012 June with the ATCA. The radial velocity axis is
  given in the rest frame defined by the optical spectroscopic
  redshift (vertical dashed line). The data have been simultaneously
  fitted by a combined continuum polynomial and a spectral line
  model. The solid black line represents the data after subtraction of
  the best-fitting (maximum-likelihood) continuum component. The red
  line represents the best-fitting spectral line model (see text for
  details). The blue line represents the best-fitting residual,
  including a velocity-axis offset for clarity.}
\label{figure:atca_results}
\end{figure*}

Here we provide a brief summary of the method used for 21\,cm spectral
line detection and parametrization. For a more detailed description
please refer to work by \cite{Allison:2012a,Allison:2012b}, and
references therein. In order to verify whether 21\,cm \mbox{H\,{\sc
    i}} absorption has been detected with significance, we estimated
the conditional probability of a spectral line model (parametrized by
a sum of spectral line components and a first-order polynomial) and
compared it to that of a continuum-only model (also parametrized by a
first-order polynomial). We also included a nuisance parameter, which
represents systematic error due to imperfect data reduction. The model
spectrum is multiplied by this parameter, with a non-uniform prior
probability given by a normal distribution of $\mu = 1.0$ and $\sigma
= 0.1$, which represents a 10\,per\,cent systematic error. Note that
we have revised our estimate of the systematic error in the ATCA data
with respect to that given by \cite{Allison:2012a}.

We have assumed no prior information about the probability of
detection of a spectral-line, since it is difficult to draw strong
statistical conclusions from the existing small sample of detected
associated \mbox{H\,{\sc i}} absorption systems
\citep[e.g.][]{Curran:2010, Curran:2011a, Allison:2012a}. Therefore,
the ratio of probabilites for each model hypothesis, given the data,
is equal to the ratio of probabilities for the data given the model
hypotheses (also known as the ratio of marginal likelihoods or
Bayesian evidence). We define the statistic $R$ by
\begin{equation}\label{equation:definition_R}
  R \equiv \ln\left({E_\mathrm{HI}\over  E_\mathrm{cont}}\right),
\end{equation}
where $E_\mathrm{HI}$ and $E_\mathrm{cont}$ are the Bayesian evidence
for the spectral line and continuum-only models respectively. If the
value of $R$ is greater than zero, then this gives the level of
significance for the spectral line hypothesis. However, if the value
of $R$ is less than zero, then the data do not warrant the inclusion
of a spectral line component in the model and so we reject this
hypothesis.

In order to calculate the Bayesian evidence, we used an efficient
Monte Carlo algorithm provided by \cite{Feroz:2008} and
\cite{Feroz:2009b}, called \textsc{MultiNest}, which implements the
nested sampling algorithm by \cite{Skilling:2004}. We optimized the
significance statistic $R$ by increasing the number of spectral line
components, thus giving the best-fitting spectral line model.

\section{Results and Discussion}

\subsection{Parametrization of the 21\,cm absorption}

We implemented the analysis method outlined in
Section\,\ref{section:analysis}, initially representing the 21\,cm
spectral line by multiple Gaussian components. The data in both epochs
warrant the inclusion of the spectral line model over the
continuum-only model and, by comparing the Bayesian evidence for
different numbers of spectral line components, the single-component
model is favoured. However, while a single Gaussian does provide a
good fit to the combined data (at maximum likelihood
$\chi^{2}_\mathrm{ml}/\mathrm{d.o.f} = 0.94\pm0.06$), upon visual
inspection the spectral line profile appears more steep sided than
predicted by this model. To test this hypothesis, we tried an
empirically motivated spectral line model, which provides for a
simultaneously broad and steep-sided trough-like profile, and is given
by the product of two Gauss error functions (R.~Jurek and
T.~Westmeier, private communication). This is a simplified version of
the 21\,cm spectral line model developed by Westmeier, Jurek \&
Obreschkow (in preparation). Since this model contains one more
parameter than a single Gaussian profile, which determines the
steepness of the profile sides, it must provide a significantly better
representation of the data in order to be preferred.

Again using Bayesian inference, we found that the 2011 April data
favour the single Gaussian model, while the 2012 June and combined
data favour the trough-like model. This result is perhaps unsurprising
given the relatively low signal-to-noise ratio (S/N) of the 2011 April
data. In the case of the combined data, the trough-like model is found
to be approximately 35 times more likely than the single Gaussian
model, with a significance above the continuum-only model of $R =
109.5 \pm 0.1$. When we replaced the uniform prior for the redshift of
the 21\,cm spectral-line with a normal prior of $12859 \pm
500$\,km\,s$^{-1}$, given by the 6dFGS optical redshift
\citep{Jones:2009} and typical infall/outflow gas velocities, the
spectral line significance increases to $R =
111.8\pm0.1$. Fig.\,\ref{figure:atca_results} shows the best-fitting
spectral line model for each of the individual spectra and the
velocity-binned combined data.

For a sufficiently bright background source, the actual optical depth
is related to the continuum flux density ($S_\mathrm{cont}$), the
fraction of flux that is covered by the foreground \mbox{H\,{\sc i}}
gas ($f$) and the observed spectral-line depth
($\Delta{S_\mathrm{line}}$). This relationship is given by
\begin{equation}
  \tau(v) = -\ln{\left(1 - {\Delta{S_\mathrm{line}} \over f S_\mathrm{cont}}\right)},
\end{equation}
where $\tau(v)$ is the optical depth of the absorbing gas as a
function of the rest-frame radial velocity. If we assume that the
21\,cm absorption is optically thin ($\Delta{S_\mathrm{line}}/f
S_\mathrm{cont}~\lesssim~0.3$), the above expression reduces to
\begin{equation}
  \tau(v) \approx {\Delta{S_\mathrm{line}} \over f S_\mathrm{cont}} \approx {\tau_\mathrm{obs}(v) \over f},
\end{equation}
where $\tau_\mathrm{obs}(v)$ is the observed optical depth. Since many
21\,cm absorption lines in the literature are far from being well
modelled by a single Gaussian component, for the purpose of
comparison, we define an effective width by
\begin{equation}\label{equation:effective_width}
  \Delta{v}_\mathrm{eff} \equiv {\int{\tau_\mathrm{obs}(v)\,\mathrm{d}v} \over \tau_\mathrm{obs,peak}},
\end{equation}
where $\tau_\mathrm{obs,peak}$ is the peak observed optical
depth. This definition of the spectral line width avoids missing the
contribution of broad, shallow components (as when using the FWHM), or
the width parameter being strongly dependent on the spectral
uncertainty (as when using the full width at zero-intensity). For the
special case of a single Gaussian component, the effective width is
just a factor of 1.06 times the FWHM. In
Table\,\ref{table:hi_summary}, we summarize the derived spectral line
parameters from the best-fitting models. Parameter values estimated
from the 2011 April and 2012 June data are mostly consistent with each
other and any significant differences are the result of the choice of
model parametrization (the single Gaussian is intrinsically more
peaked than the trough-like model). The small difference between the
value of $R$ for our re-analysis of the 2011 April data, with that
from the previous analysis by \cite{Allison:2012a}, is the result of
improvements to our data reduction method.

\begin{table*}
  \centering
  \caption{A summary of the derived properties of \mbox{H\,{\sc i}} absorption
    from model fitting to the ATCA 21\,cm data, where
    $\sigma_\mathrm{chan}$ is the estimated uncertainty per channel;
    $cz_\mathrm{peak}$ is the redshift at peak 
    spectral-line depth; $\Delta{S}_\mathrm{line,peak}$ is the peak spectral line depth; 
    ${S}_\mathrm{cont,peak}$ is the continuum flux at peak spectral line depth;
    $\int{\tau_\mathrm{obs}(v)\mathrm{d}v}$ is the 
    velocity integrated observed optical depth; $\Delta{v}_\mathrm{eff}$ is the effective 
    velocity width (as defined by equation\,\ref{equation:effective_width}); 
    $R$ is the natural logarithm of the ratio of
    probabilities for the single-component spectral line model versus
    the continuum-only model; $\chi^{2}_\mathrm{ml}/\mathrm{d.o.f}$
    is the reduced chi-squared statistic for the best-fitting
    (maximum likelihood) model parameters (under the assumption of
    model linearity). Note that the uncertainties include the
    systematic error in the data represented by a nuisance parameter
    with a normal prior.}\label{table:hi_summary}
    \begin{tabular}{lccccccccc}
      \hline
      Epoch & Resolution & $\sigma_\mathrm{chan}$ & $cz_\mathrm{peak}$ & $\Delta{S}_\mathrm{line,peak}$ & $S_\mathrm{cont,peak}$ & $\int{\tau_\mathrm{obs}(v)\mathrm{d}v}$ & $\Delta{v}_\mathrm{eff}$ & $R$ & $\chi^{2}_\mathrm{ml}/\mathrm{d.o.f}$ \\
      & (km\,s$^{-1}$) & (mJy) & (km\,s$^{-1}$) & (mJy) & (mJy) & (km\,s$^{-1}$) & (km\,s$^{-1}$) & & \\
      \hline
      2011 April & 7 & 3.83 & $13073\pm38$ & $4.8\pm0.8$ & $160\pm11$ & $15.8\pm2.8$ & $540\pm120$ & $24.2\pm0.1$ & $0.98\pm0.03$ \\
      2012 June & 7 & 1.91 & $13048\pm10$ & $3.9\pm0.4$ & $163\pm10$ & $10.1\pm0.7$ & $418\pm20$ & $95.0\pm0.1$ & $1.02\pm0.03$ \\
      Combined & 21 & 1.05 & $13046\pm10$ & $4.1\pm0.4$ & $163\pm10$ & $10.5\pm0.7$ & $418\pm20$ & $109.5\pm0.1$ & $0.91\pm0.06$ \\
      \hline
    \end{tabular}
\end{table*}

The column density of atomic neutral hydrogen (in cm$^{-2}$) as a
function of the integrated optical depth (in km\,s$^{-1}$) is given by
\citep{Wolfe:1975}
\begin{eqnarray}
  N_\mathrm{HI} & = & 1.823 \times 10^{18}\,T_\mathrm{spin} \int{\tau(v)\,\mathrm{d}v} \nonumber \\
  & \approx & 1.823 \times 10^{18}\,{T_\mathrm{spin} \over f} \int{\tau_\mathrm{obs}(v)\,\mathrm{d}v},
\end{eqnarray}
where $T_\mathrm{spin}$ is the mean harmonic spin temperature of the
gas (in K). Based on this relation, we estimate the \mbox{H\,{\sc i}}
column density towards PMN\,J2054--4242 to be
\begin{equation}
  N_\mathrm{HI} \approx 1.91 \pm 0.14 \times 10^{21} \left({T_\mathrm{spin}\over 100\,\mathrm{K}}\right) \left({1.0\over f}\right)\,\mathrm{cm}^{-2},
\end{equation}
which for values of $T_\mathrm{spin}/f \gtrsim 10\,$K is equivalent to
that of a damped Lyman-$\alpha$ absorber\footnote{The lowest value of
  $T_\mathrm{spin}/f$ found so far, for a damped Lyman-$\alpha$
  absorber, is 60\,K \citep{Curran:2007}.} ($N_\mathrm{HI} > 2 \times
10^{20}\,\mathrm{cm}^{-2}$).

\subsection{High velocity cold gas towards PMN\,J2054--4242}

By applying Bayesian inference to our 21\,cm ATCA data we have
detected cold, high column density \mbox{H\,{\sc i}} gas towards the
compact flat-spectrum radio source PMN\,J2054--4242. The observed
total integrated optical depth for the absorption is relatively large
(see Table\,\ref{table:hi_summary}), however, this is distributed over
a broad spectral line width, producing a modest peak optical depth of
$\tau_\mathrm{obs,peak} = 2.5\pm0.2$\,per\,cent. The 21\,cm spectral
line is redshifted with respect to the optical spectroscopic redshift
\citep[$z_\mathrm{opt} = 0.042893\pm0.000150$;][]{Jones:2009} by
$187\pm46$\,km\,s$^{-1}$, possibly indicating that the gas is
infalling towards the central radio source. The strong broadening of
the line perhaps suggests that the cold gas is rotating with very high
velocity, or that the infall is accelerated along the line of sight. A
high spatial resolution 21\,cm study of the similarly compact radio
source PKS\,B1814--637, by \citet{Morganti:2011}, showed that the
broadened absorption component in that system was likely located
within a circumnuclear disc, while the deeper narrow component arises
from absorption in an extended disc of gas on the kpc scale. The
absence of a narrow component in the spectrum of PMN\,J2054--4242 is
perhaps indicative of either low column density (or missing)
large-scale cold gas, or that any extended structure is orientated
away from the line of sight to the radio source. These hypotheses are
consistent with the optical classifications of the host galaxies for
each of these sources: PKS\,B1814--637 is hosted by a gas-rich regular
edge-on disc, while for PMN\,J2054-4242 the host is an early type
\citep{Loveday:1996}. However, without further information, we can
only speculate here as to the physical nature of the spectral line
broadening. We are therefore pursuing follow-up multiwavelength and
high spatial resolution observations to test these interpretations of
the 21\,cm ATCA data.

\subsection{The scarcity of broad, weak 21\,cm absorption lines}

The left-hand panel of Fig.\,\ref{figure:atca_results} shows the
initial detection of broad 21\,cm absorption towards PMN\,J2054--4242
(2011 April). From visual inspection of this spectrum, one might
reject the feature as being an artefact, due to its wide profile and
low peak S/N. Indeed by using the automated source-finding tool
\textsc{Duchamp} \citep{Whiting:2012}, this original feature was only
detected with significance once we had smoothed the spectral data
(using a Hann window function with a width of 5) and applied a
threshold cut-off of three times the S/N. However, by using a Bayesian
approach to spectral line finding and fitting, \citet{Allison:2012a}
found that the feature was significant above the noise and that it
occupied a parameter space that was different to that of the false
detections from continuum artefacts. The redshift is close to that of
the optical spectroscopic measurement, which strongly suggests that it
arises from associated \mbox{H\,{\sc i}} absorption. All of these
indicators prompted further observations in 2012 June, with a longer
integration time, leading to confirmation that the feature is indeed a
21\,cm absorpion line.

Such broad, shallow 21\,cm absorption lines physically correspond to a
large spread in the velocity of \mbox{H\,{\sc i}} gas along the line
of sight. Existing detections of broad absorption are often associated
with an acompanying narrow, deeper component. For example,
\cite{Morganti:2005b} conducted a targeted survey of 11 bright
radio-loud AGNs, with evidence of past star formation and a rich
inter-stellar medium, and most with existing detections of deep,
narrow \mbox{H\,{\sc i}} absorption. The broad bandwidth available
($\pm 2000$\,km\,s$^{-1}$) and the high sensitivity achieved (rms
$\sim$\,0.4 -- 0.8\,mJy\,beam$^{-1}$), using the Westerbork Synthesis
Radio Telescope, allowed detection of broad and shallower absorption
with $\tau_\mathrm{peak} \lesssim 0.01$. Morganti et al. detected
broad, shallow absorption towards 6 radio galaxies, all of which were
associated with existing deep, narrow components, and had bulk
blueshifted motion interpreted as line-of-sight cold gas
outflows. However, our detection of a single broad absorption line
towards PMN\,J2054-4242 raises the question as to whether there exist
a population of very shallow and broad \mbox{H\,{\sc i}} absorption
lines, which have no associated narrow, deep component, and as yet
remain largely undetected.

\begin{figure}
\centering
\includegraphics[width = 1.0\columnwidth]{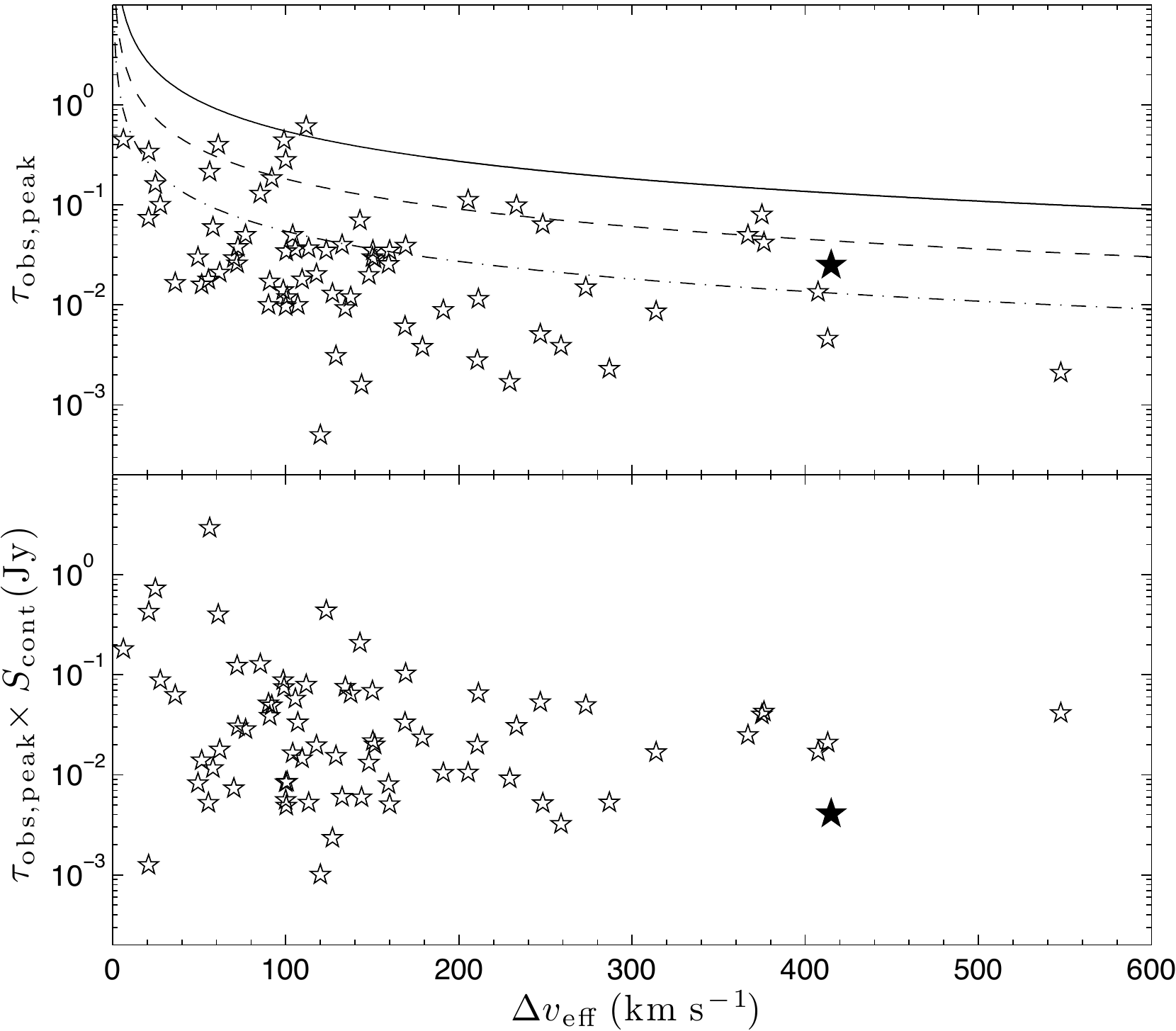}
\caption{The observed 21\,cm peak optical depth (top), and in flux
  units (bottom), versus the spectral line width (as defined by
  equation\,\ref{equation:effective_width}), for detections of
  associated \mbox{H\,{\sc i}} absorption. The empty stars represent a
  sample from the literature (see text for details and references
  therein) and the filled star represents our detection towards
  PMN\,J2054--4242. The lines represent \mbox{H\,{\sc i}} absorption
  for the maximum observed column density
  $\sim$\,10$^{22}$\,cm$^{-2}$, and constant $T_\mathrm{spin}/f$ of
  100\,K (solid), 300\,K (dashed) and 1000\,K (dot-dashed).}
\label{figure:literature_widths}
\end{figure}

To address this question, we show in the top panel of
Fig.\,\ref{figure:literature_widths} the peak observed optical depth,
versus the effective spectral line width, for a representative sample
of detections of 21\,cm associated \mbox{H\,{\sc i}} absorption in the
literature\footnote{References for the literature sample shown in
  Fig.\ref{figure:literature_widths}: \citet{Shostak:1983},
  \citet{VanGorkom:1986,VanGorkom:1989},
  \citet{Mirabel:1989,Mirabel:1990}, \citet{Jaffe:1990},
  \citet{Uson:1991}, \citet{Carilli:1992,Carilli:1998},
  \citet{Morganti:1998}, \citet{Moore:1999},
  \citet{Peck:1999,Peck:2000}, \citet{Morganti:2001},
  \citet{Ishwara-Chandra:2003}, \citet{Vermeulen:2003},
  \citet{Maness:2004}, \citet{Morganti:2005a}, \citet{Gupta:2006b},
  \citet{Curran:2006}, \citet{Orienti:2006}, \citet{Gupta:2006a},
  \citet{Emonts:2008}, \citet{Struve:2010}, \citet{Salter:2010},
  \citet{Emonts:2010}, \citet{Curran:2011a,Curran:2011b},
  \citet{Chandola:2011}, \citet{Emonts:2012} and
  \citet{Allison:2012a}.}  (\citealt{Curran:2013} also present a
similar plot of the FWHM for a literature sample of redshifted
absorbers). We have treated multiple spectral components as a single
detection, so that a profile exhibiting a single broad component will
have a larger effective width than that with both broad and strong
narrow components. Upon inspection of
Fig.\,\ref{figure:literature_widths}, it is evident that these data
exhibit a weak anticorrelation between the peak observed optical depth
and spectral line width (the Pearson product moment correlation
coefficient $r \approx -0.3$), consistent with the maximum column
density limit $\sim 10^{22}$\,cm$^{-2}$, seen in observations of
\mbox{H\,{\sc i}} gas both at $z \approx 0$ \citep[][]{Zwaan:2005} and
$z \approx 3$ \citep[][]{Altay:2011}. However, it should be noted that
this figure is purely indicative of the overall trend, since we have
not accounted for the myriad uncertainties in these individual
estimates.

The large majority of detections in the literature sample have
spectral line widths smaller than our detection towards
PMN\,J2054--4242 making this one of the broadest 21\,cm absorption
lines yet detected. The widest spectral line shown in
Fig.\,\ref{figure:literature_widths} arises from absorption in the
central galaxy NGC\,1275 of the Perseus Cluster
\citep[$\sim$\,550\,km\,s$^{-1}$;][]{Jaffe:1990}, which similarly has
a profile dominated by broad absorption. However, the background radio
source in this detection (3C\,84) has a flux density of 19.7\,Jy,
making it a substantially more powerful probe of low optical depth gas
than PMN\,J2054-4242. Therefore, in the bottom panel of
Fig.\,\ref{figure:literature_widths}, we have plotted the product of
the peak observed optical depth and continuum flux density versus the
spectral line width. Our detection occupies a region of parameter
space with low spectral line depth and broad width, which has not yet
been largely explored by existing searches for associated
\mbox{H\,{\sc i}} absorption. Finding such a rarely detected system,
in data with a relatively low S/N, demonstrates the power of our
analysis method. The development of such sophisticated automated
spectral line finding techniques will be vitally important for
detection of these broad, weak absorption lines in future large-scale
21\,cm absorption surveys.

\section*{Acknowledgments} 
We thank Russell Jurek and Tobias Westmeier for useful discussions on
21\,cm spectral-line parametrization. JRA acknowledges support from an
ARC Super Science Fellowship. The Centre for All-sky Astrophysics is
an Australian Research Council Centre of Excellence, funded by grant
CE110001020. The ATCA is part of the Australia Telescope National
Facility which is funded by the Commonwealth of Australia for
operation as a National Facility managed by CSIRO. Computing
facilities were provided by the High Performance Computing Facility at
the University of Sydney. This research has made use of the NASA/IPAC
Extragalactic Database (NED) which is operated by the Jet Propulsion
Laboratory, California Institute of Technology, under contract with
the National Aeronautics and Space Administration. This research has
also made use of NASA's Astrophysics Data System Bibliographic
Services.

\bibliographystyle{mn2e.bst}
\bibliography{./bibliography.bib}

\begin{thebibliography}{51}
\expandafter\ifx\csname natexlab\endcsname\relax\def\natexlab#1{#1}\fi

\bibitem[{{Allison} {et~al}\mbox{.}(2012{\natexlab{a}}){Allison}, {Curran},
  {Emonts}, {Ger{\'e}b}, {Mahony}, {Reeves}, {Sadler}, {Tanna}, {Whiting}, \&
  {Zwaan}}]{Allison:2012a}
{Allison} J.~R. {et~al.}, 2012{\natexlab{a}}, MNRAS, 423, 2601

\bibitem[{{Allison} {et~al}\mbox{.}(2012{\natexlab{b}}){Allison}, {Sadler}, \&
  {Whiting}}]{Allison:2012b}
{Allison} J.~R., {Sadler} E.~M., {Whiting} M.~T., 2012{\natexlab{b}}, PASA, 29,
  221

\bibitem[{{Altay} {et~al}\mbox{.}(2011){Altay}, {Theuns}, {Schaye}, {Crighton},
  \& {Dalla Vecchia}}]{Altay:2011}
{Altay} G., {Theuns} T., {Schaye} J., {Crighton} N.~H.~M., {Dalla Vecchia} C.,
  2011, ApJL, 737, L37

\bibitem[{{Carilli} {et~al}\mbox{.}(1998){Carilli}, {Menten}, {Reid}, {Rupen},
  \& {Yun}}]{Carilli:1998}
{Carilli} C.~L., {Menten} K.~M., {Reid} M.~J., {Rupen} M.~P., {Yun} M.~S.,
  1998, ApJ, 494, 175

\bibitem[{{Carilli} {et~al}\mbox{.}(1992){Carilli}, {Perlman}, \&
  {Stocke}}]{Carilli:1992}
{Carilli} C.~L., {Perlman} E.~S., {Stocke} J.~T., 1992, ApJL, 400, L13

\bibitem[{{Chandola} {et~al}\mbox{.}(2011){Chandola}, {Sirothia}, \&
  {Saikia}}]{Chandola:2011}
{Chandola} Y., {Sirothia} S.~K., {Saikia} D.~J., 2011, MNRAS, 418, 1787

\bibitem[{{Curran} {et~al}\mbox{.}(2007){Curran}, {Tzanavaris}, {Murphy},
  {Webb}, \& {Pihlstr{\"o}m}}]{Curran:2007}
{Curran} S.~J., {Tzanavaris} P., {Murphy} M.~T., {Webb} J.~K., {Pihlstr{\"o}m}
  Y.~M., 2007, MNRAS, 381, L6

\bibitem[{{Curran} \& {Whiting}(2010)}]{Curran:2010}
{Curran} S.~J., {Whiting} M.~T., 2010, ApJ, 712, 303

\bibitem[{{Curran} {et~al}\mbox{.}(2011{\natexlab{a}}){Curran}, {Whiting},
  {Murphy}, {Webb}, {Bignell}, {Polatidis}, {Wiklind}, {Francis}, \&
  {Langston}}]{Curran:2011a}
{Curran} S.~J. {et~al.}, 2011{\natexlab{a}}, MNRAS, 413, 1165

\bibitem[{{Curran} {et~al}\mbox{.}(2006){Curran}, {Whiting}, {Murphy}, {Webb},
  {Longmore}, {Pihlstr{\"o}m}, {Athreya}, \& {Blake}}]{Curran:2006}
{Curran} S.~J., {Whiting} M.~T., {Murphy} M.~T., {Webb} J.~K., {Longmore}
  S.~N., {Pihlstr{\"o}m} Y.~M., {Athreya} R., {Blake} C., 2006, MNRAS, 371, 431

\bibitem[{{Curran} {et~al}\mbox{.}(2013){Curran}, {Whiting}, {Sadler}, \&
  {Bignell}}]{Curran:2013}
{Curran} S.~J., {Whiting} M.~T., {Sadler} E.~M., {Bignell} C., 2013, MNRAS,
  428, 2053

\bibitem[{{Curran} {et~al}\mbox{.}(2011{\natexlab{b}}){Curran}, {Whiting},
  {Webb}, \& {Athreya}}]{Curran:2011b}
{Curran} S.~J., {Whiting} M.~T., {Webb} J.~K., {Athreya} R.,
  2011{\natexlab{b}}, MNRAS, 414, L26

\bibitem[{{Emonts} {et~al}\mbox{.}(2012){Emonts}, {Burnett}, {Morganti}, \&
  {Struve}}]{Emonts:2012}
{Emonts} B.~H.~C., {Burnett} C., {Morganti} R., {Struve} C., 2012, MNRAS, 421,
  1421

\bibitem[{{Emonts} {et~al}\mbox{.}(2008){Emonts}, {Morganti}, {Oosterloo},
  {Holt}, {Tadhunter}, {van der Hulst}, {Ojha}, \& {Sadler}}]{Emonts:2008}
{Emonts} B.~H.~C., {Morganti} R., {Oosterloo} T.~A., {Holt} J., {Tadhunter}
  C.~N., {van der Hulst} J.~M., {Ojha} R., {Sadler} E.~M., 2008, MNRAS, 387,
  197

\bibitem[{{Emonts} {et~al}\mbox{.}(2010){Emonts}, {Morganti}, {Struve},
  {Oosterloo}, {van Moorsel}, {Tadhunter}, {van der Hulst}, {Brogt}, {Holt}, \&
  {Mirabal}}]{Emonts:2010}
{Emonts} B.~H.~C. {et~al.}, 2010, MNRAS, 406, 987

\bibitem[{{Feroz} \& {Hobson}(2008)}]{Feroz:2008}
{Feroz} F., {Hobson} M.~P., 2008, MNRAS, 384, 449

\bibitem[{{Feroz} {et~al}\mbox{.}(2009){Feroz}, {Hobson}, \&
  {Bridges}}]{Feroz:2009b}
{Feroz} F., {Hobson} M.~P., {Bridges} M., 2009, MNRAS, 398, 1601

\bibitem[{{Griffith} \& {Wright}(1993)}]{Griffith:1993}
{Griffith} M.~R., {Wright} A.~E., 1993, AJ, 105, 1666

\bibitem[{{Gupta} \& {Saikia}(2006)}]{Gupta:2006b}
{Gupta} N., {Saikia} D.~J., 2006, MNRAS, 370, L80

\bibitem[{{Gupta} {et~al}\mbox{.}(2006){Gupta}, {Salter}, {Saikia}, {Ghosh}, \&
  {Jeyakumar}}]{Gupta:2006a}
{Gupta} N., {Salter} C.~J., {Saikia} D.~J., {Ghosh} T., {Jeyakumar} S., 2006,
  MNRAS, 373, 972

\bibitem[{{Healey} {et~al}\mbox{.}(2007){Healey}, {Romani}, {Taylor}, {Sadler},
  {Ricci}, {Murphy}, {Ulvestad}, \& {Winn}}]{Healey:2007}
{Healey} S.~E., {Romani} R.~W., {Taylor} G.~B., {Sadler} E.~M., {Ricci} R.,
  {Murphy} T., {Ulvestad} J.~S., {Winn} J.~N., 2007, ApJS, 171, 61

\bibitem[{{Ishwara-Chandra} {et~al}\mbox{.}(2003){Ishwara-Chandra},
  {Dwarakanath}, \& {Anantharamaiah}}]{Ishwara-Chandra:2003}
{Ishwara-Chandra} C.~H., {Dwarakanath} K.~S., {Anantharamaiah} K.~R., 2003, J.
  Astrophys. Astron., 24, 37

\bibitem[{{Jaffe}(1990)}]{Jaffe:1990}
{Jaffe} W., 1990, A\&A, 240, 254

\bibitem[{{Jones} {et~al}\mbox{.}(2009){Jones}, {Read}, {Saunders}, {Colless},
  {Jarrett}, {Parker}, {Fairall}, {Mauch}, {Sadler}, {Watson}, {Burton},
  {Campbell}, {Cass}, {Croom}, \& et~al.}]{Jones:2009}
{Jones} D.~H. {et~al.}, 2009, MNRAS, 399, 683

\bibitem[{{Loveday}(1996)}]{Loveday:1996}
{Loveday} J., 1996, MNRAS, 278, 1025

\bibitem[{{Maness} {et~al}\mbox{.}(2004){Maness}, {Taylor}, {Zavala}, {Peck},
  \& {Pollack}}]{Maness:2004}
{Maness} H.~L., {Taylor} G.~B., {Zavala} R.~T., {Peck} A.~B., {Pollack} L.~K.,
  2004, ApJ, 602, 123

\bibitem[{{Mirabel}(1989)}]{Mirabel:1989}
{Mirabel} I.~F., 1989, ApJL, 340, L13

\bibitem[{{Mirabel}(1990)}]{Mirabel:1990}
{Mirabel} I.~F., 1990, ApJL, 352, L37

\bibitem[{{Moore} {et~al}\mbox{.}(1999){Moore}, {Carilli}, \&
  {Menten}}]{Moore:1999}
{Moore} C.~B., {Carilli} C.~L., {Menten} K.~M., 1999, ApJL, 510, L87

\bibitem[{{Morganti} {et~al}\mbox{.}(2009){Morganti}, {Emonts}, \&
  {Oosterloo}}]{Morganti:2009a}
{Morganti} R., {Emonts} B., {Oosterloo} T., 2009, A\&A, 496, L9

\bibitem[{{Morganti} {et~al}\mbox{.}(2011){Morganti}, {Holt}, {Tadhunter},
  {Ramos Almeida}, {Dicken}, {Inskip}, {Oosterloo}, \&
  {Tzioumis}}]{Morganti:2011}
{Morganti} R., {Holt} J., {Tadhunter} C., {Ramos Almeida} C., {Dicken} D.,
  {Inskip} K., {Oosterloo} T., {Tzioumis} T., 2011, A\&A, 535, A97

\bibitem[{{Morganti} {et~al}\mbox{.}(1998){Morganti}, {Oosterloo}, \&
  {Tsvetanov}}]{Morganti:1998}
{Morganti} R., {Oosterloo} T., {Tsvetanov} Z., 1998, AJ, 115, 915

\bibitem[{{Morganti} {et~al}\mbox{.}(2005{\natexlab{a}}){Morganti},
  {Oosterloo}, {Tadhunter}, {van Moorsel}, \& {Emonts}}]{Morganti:2005a}
{Morganti} R., {Oosterloo} T.~A., {Tadhunter} C.~N., {van Moorsel} G., {Emonts}
  B., 2005{\natexlab{a}}, A\&A, 439, 521

\bibitem[{{Morganti} {et~al}\mbox{.}(2001){Morganti}, {Oosterloo}, {Tadhunter},
  {van Moorsel}, {Killeen}, \& {Wills}}]{Morganti:2001}
{Morganti} R., {Oosterloo} T.~A., {Tadhunter} C.~N., {van Moorsel} G.,
  {Killeen} N., {Wills} K.~A., 2001, MNRAS, 323, 331

\bibitem[{{Morganti} {et~al}\mbox{.}(2005{\natexlab{b}}){Morganti},
  {Tadhunter}, \& {Oosterloo}}]{Morganti:2005b}
{Morganti} R., {Tadhunter} C.~N., {Oosterloo} T.~A., 2005{\natexlab{b}}, A\&A,
  444, L9

\bibitem[{{Orienti} {et~al}\mbox{.}(2006){Orienti}, {Morganti}, \&
  {Dallacasa}}]{Orienti:2006}
{Orienti} M., {Morganti} R., {Dallacasa} D., 2006, A\&A, 457, 531

\bibitem[{{Peck} {et~al}\mbox{.}(1999){Peck}, {Taylor}, \&
  {Conway}}]{Peck:1999}
{Peck} A.~B., {Taylor} G.~B., {Conway} J.~E., 1999, ApJ, 521, 103

\bibitem[{{Peck} {et~al}\mbox{.}(2000){Peck}, {Taylor}, {Fassnacht},
  {Readhead}, \& {Vermeulen}}]{Peck:2000}
{Peck} A.~B., {Taylor} G.~B., {Fassnacht} C.~D., {Readhead} A.~C.~S.,
  {Vermeulen} R.~C., 2000, ApJ, 534, 104

\bibitem[{{Salter} {et~al}\mbox{.}(2010){Salter}, {Saikia}, {Minchin}, {Ghosh},
  \& {Chandola}}]{Salter:2010}
{Salter} C.~J., {Saikia} D.~J., {Minchin} R., {Ghosh} T., {Chandola} Y., 2010,
  ApJL, 715, L117

\bibitem[{{Sault} {et~al}\mbox{.}(1995){Sault}, {Teuben}, \&
  {Wright}}]{Sault:1995}
{Sault} R.~J., {Teuben} P.~J., {Wright} M.~C.~H., 1995, in ASP Conf. Ser.,
  Vol.~77, Astronomical Data Analysis Software and Systems IV, {Shaw R.~A.,
  Payne H.~E., \& Hayes J.~J.~E.}, ed., Astron. Soc. Pac., San Francisco, p.
  433

\bibitem[{{Shostak} {et~al}\mbox{.}(1983){Shostak}, {van Gorkom}, {Ekers},
  {Sanders}, {Goss}, \& {Cornwell}}]{Shostak:1983}
{Shostak} G.~S., {van Gorkom} J.~H., {Ekers} R.~D., {Sanders} R.~H., {Goss}
  W.~M., {Cornwell} T.~J., 1983, A\&A, 119, L3

\bibitem[{{Skilling}(2004)}]{Skilling:2004}
{Skilling} J., 2004, in {AIP Conf. Ser.}, Vol. 735, Bayesian Inference and
  Maximum Entropy methods in Science and Engineering, {Fischer R., Preuss R. \&
  Toussaint U. V.}, ed., {Am. Inst. Phys.}, New York, p. 395

\bibitem[{{Struve} \& {Conway}(2010)}]{Struve:2010}
{Struve} C., {Conway} J.~E., 2010, A\&A, 513, A10

\bibitem[{{Uson} {et~al}\mbox{.}(1991){Uson}, {Bagri}, \&
  {Cornwell}}]{Uson:1991}
{Uson} J.~M., {Bagri} D.~S., {Cornwell} T.~J., 1991, Phys. Rev. Lett., 67, 3328

\bibitem[{{van Gorkom} {et~al}\mbox{.}(1989){van Gorkom}, {Knapp}, {Ekers},
  {Ekers}, {Laing}, \& {Polk}}]{VanGorkom:1989}
{van Gorkom} J.~H., {Knapp} G.~R., {Ekers} R.~D., {Ekers} D.~D., {Laing} R.~A.,
  {Polk} K.~S., 1989, AJ, 97, 708

\bibitem[{{van Gorkom} {et~al}\mbox{.}(1986){van Gorkom}, {Knapp}, {Raimond},
  {Faber}, \& {Gallagher}}]{VanGorkom:1986}
{van Gorkom} J.~H., {Knapp} G.~R., {Raimond} E., {Faber} S.~M., {Gallagher}
  J.~S., 1986, AJ, 91, 791

\bibitem[{{Vermeulen} {et~al}\mbox{.}(2003){Vermeulen}, {Pihlstr{\"o}m},
  {Tschager}, {de Vries}, {Conway}, {Barthel}, {Baum}, {Braun}, {Bremer},
  {Miley}, {O'Dea}, {R{\"o}ttgering}, {Schilizzi}, {Snellen}, \&
  {Taylor}}]{Vermeulen:2003}
{Vermeulen} R.~C. {et~al.}, 2003, A\&A, 404, 861

\bibitem[{{Whiting}(2012)}]{Whiting:2012}
{Whiting} M.~T., 2012, MNRAS, 421, 3242

\bibitem[{{Wilson} {et~al}\mbox{.}(2011){Wilson}, {Ferris}, {Axtens}, {Brown},
  {Davis}, {Hampson}, {Leach}, {Roberts}, {Saunders}, {Koribalski}, {Caswell},
  {Lenc}, {Stevens}, {Voronkov}, \& et~al.}]{Wilson:2011}
{Wilson} W.~E. {et~al.}, 2011, MNRAS, 416, 832

\bibitem[{{Wolfe} \& {Burbidge}(1975)}]{Wolfe:1975}
{Wolfe} A.~M., {Burbidge} G.~R., 1975, ApJ, 200, 548

\bibitem[{{Zwaan} {et~al}\mbox{.}(2005){Zwaan}, {van der Hulst}, {Briggs},
  {Verheijen}, \& {Ryan-Weber}}]{Zwaan:2005}
{Zwaan} M.~A., {van der Hulst} J.~M., {Briggs} F.~H., {Verheijen} M.~A.~W.,
  {Ryan-Weber} E.~V., 2005, MNRAS, 364, 1467

\end{thebibliography}

\bsp

\label{lastpage}

\end{document}